# Quantum Computing without Quantum Computers: Database Search and Data Processing Using Classical Wave Superposition

Michael  Balynskiy[1], Howard Chiang [1], David Gutierrez[1], Alexander Kozhevnikov[2], Yuri Filimonov[2,3], and  Alexander Khitun[1*]

[1]*Department of Electrical and Computer Engineering, University of California -Riverside, Riverside, California, USA 92521*

[2]*Kotelnikov Institute of Radioengineering and Electronics of the Russian Academy of Sciences, Saratov, Russia 410019*

[3]*Saratov State University, Saratov, Russia 410012*

Quantum computing is an emerging field  of science which will eventually lead us to new and powerful logic devices with capabilities far beyond the limits of current transistor-based technology  [1]. There are certain types of problems which quantum computers can solve fundamentally faster than the tradition digital computers. For example,  Peter Shor developed quantum algorithm which solves the factoring and discrete logarithm problems in time $O(n^3)$, compared with the exponential time  required for the best known classical algorithm [2]. Quantum computers can search an "unsorted database" (that is, for $f(x):\{0,N\} \rightarrow \{0,1\}$, find $x_0$ such that $f(x_0) = 1$)in time $O(\sqrt{N})$ , compared with the $O(N)$  time that would be required classically [3]. Superposition of states and entanglement are the key two ingredients which make quantum computing so powerful.  Superposition of states allows us to speedup database search by checking a number of bits in parallel, while quantum entanglement is critically important for quantum cryptography [4]. There are quantum algorithms which require both superposition and entanglement (e.g. Shor's algorithm). But neither the Grover algorithm nor the very first quantum algorithm due to Deutsch and Jozsa [5] need entanglement [6]. *Is it possible to utilize classical wave superposition to speedup database search?* This interesting question was analyzed by S. Lloyd [7]. It was concluded that classical devices that rely on *wave interference may provide the same speedup over classical digital devices as quantum devices* [7]. A. Patel came up with the same conclusion on the efficiency of using classical wave superposition for database search [8].  There were several experimental works using optical beam superposition for emulating Grover's algorithm [9-11]. It was concluded that the use of classical wave superposition comes with the cost of exponential increase of the resources [9]. Since then, it is widely believed that the use of classical wave superposition for quantum algorithms is inevitably leading to an exponential resources overhead (e.g., number of devices, power consumption, precision requirements). In this work, we describe a classical Oracle machine which utilizes classical wave superposition for database search and data processing. We present experimental data on magnetic database search using spin wave superposition. The data show a fundamental speedup over the digital computers without any exponential resource overhead. We argue that in some cases the classical wave-based approach may provide the same speedup in database search as quantum computers. There are also examples of numerical modeling demonstrating classical wave interference for period finding.  This approach may be not compete with quantum computers with efficiency but outperform classical digital computers. There is a lot of room for classical wave-based computers development which may provide a fundamental advantage over the classical digital devices. These classical wave-based devices can perform some of the algorithms with the same efficiency as quantum computers as long as quantum entanglement is not required. The rest of the work is organized as follows. In the next Section, we describe the structure of classical Oracle and its principle of operation. Next, we present several examples of database search and period funding. The advantages and limits of the classical wave computing are provided in the Discussion Section.

Let us consider a classical Oracle machine - Oracle-C as illustrated in Fig.1. It consists of an interferometer, a non-linear detector, and a general-purpose digital computer. The interferometer is a multi-terminal device which has $n$ inputs ports and just one output port. The input and the output signals are classical sinusoidal waves (e.g., electrical, optical, sound, etc.) given by

$$y(t) = A_i \cdot \sin(\omega t + \phi_i),$$ (1)

where $A_i$ is the amplitude, $\omega$ is the frequency, $t$ is the time, and $\phi_i$ is the phase. The subscript $i$ defines the input number. All input waves have the same frequency $\omega$ and amplitude $A_0$. Information is encoded into the phases of the input waves $\phi_i$ as follows,

$$Input\ logic\ state\ (i-th\ input) = \begin{cases} 1, & if\ \ \phi_i = \pi/2 \\ 0, & if\ \ \phi_i = 0\pi \end{cases}.$$ (2)

The output is a result of $n$ wave interference:

$$A_{out} \cdot \sin(\omega t + \phi_{out}) = \sum_{i=1}^{n} \sigma_i A_i \cdot \sin(\omega t + \phi_i + \Delta_i),$$ (3)

where $\sigma_i$ is the amplitude change, $\Delta_i$ is the accumulated phase shift during wave propagation from the $i$-th input port to the output port. Hereafter, we consider a linear wave propagation (i.e., $\sigma_i$ and $\Delta_i$ do not depend on the wave amplitude). The non-linear detector detects the time-averaged power $P_{out}$ and compares it with some reference value $P_{ref}$. The output logic state is 1, if $P_{out}$ is equal or exceeds $P_{ref}$, and 0 otherwise.

$$Output\ logic\ state(output) = \begin{cases} 1, & if\ P_{out} \geq P_{ref} \\ 0, & if\ P_{out} < P_{ref} \end{cases}$$ (4)

The digital computer is aimed to control the input phases $\phi_i$, store the results of measurements, and control the reference value $P_{ref}$. *Oracle-C provides a deterministic input-output correlation*. It constitutes a database $f(x): \{0, n\} \rightarrow \{0,1\}$, where the correlation between the input and the output logic states is defined by the internal structure of the interferometer (i.e., $\sigma_i$ and $\Delta_i$). Our objective is to show that using classical wave superposition it is possible to speedup database search similar to the algorithms developed for quantum computers. In our case, the superposition of logic 0 and 1 is a wave with phase $\pi/4$:

$$\alpha|0\rangle + \beta|1\rangle \equiv \sin\phi|0\rangle + \cos\phi|1\rangle$$

$$\sin(\omega t) + \sin(\omega t + \pi/2) = \sqrt{2}\sin(\omega t + \pi/4). \qquad (5)$$

Next, we will present several examples of using Oracle-C. In Example 1, there will be shown a search through an "unsorted database", where only one input phase combination lead to logic output 1. In Example 2, the search algorithm will be extended to a database with multivalued inputs. In Example 3, we present experimental data on search through the magnetic database using spin wave superposition. Finally, in Example 4, we will present the results of numerical modeling showing period finding of a given function using Oracle–C.

## Example 1: Search through an "unsorted database"

Let us consider an Oracle-C constructed such a way that only one input combination results in the output logic state 1: $f(x_0) \rightarrow \{0,1\}$, where $x_0$ is an input phase combination $x_0 \equiv \{\phi_1, \phi_2, \dots \phi_n\}$. For instance, there is only one input phase combination resulting in the constructing wave interference at the output (i.e., all waves are coming in phase to the output). All other input phase combinations result in a lower output power. The task is to find this phase combination in the minimum number of steps. There are several assumptions we make regarding the internal interferometer structure. (i) All input waves have the same amplitude $A_0$. (ii) The amplitudes do not change during wave propagation within the interferometer (i.e., $\sigma = 1$ for all inputs). (iii) The phase shifters $\Delta_i$ may provide either $+\pi/4$ or $-\pi/4$ phase shift. There are $2^n$ possible phase combinations. There are $2^n$ ways to choose a set of $\Delta_i$. However, there are $2^n - 2$ shifter combinations leading to the only one constructive interference output. Two combinations with all $\Delta_i = +\pi/4 \ or -\pi/4$ should be excluded from consideration. The reference value in the detector is setup to $P_{ref} = n^2 P_0$, where $P_0$ is the power provided by just one input. Thus, there is *only one input phase* combination which results in the constructive wave interference providing logic output 1 for the given set of $\Delta_i$. It would take $2^{n-1}$ queries in average to accomplish this task taking one by one all possible input combinations. It takes only $2n$ queries using classical wave superposition. The solving procedure is the following.

In step 1, we find the right phase for input 1. All inputs from $i = 2$ to $i = n$ are setup in a superposition of states (i.e., by exciting input waves with a phase $\pi/4$). First, we measure the output for input combination $\{0\pi, \pi/4, \pi/4, \dots \pi/4\}$. This input phase combination is equivalent to logic state $\left\{0, \frac{|0\rangle + |1\rangle}{\sqrt{2}}, \frac{|0\rangle + |1\rangle}{\sqrt{2}}, \dots \frac{|0\rangle + |1\rangle}{\sqrt{2}}\right\}$. The output $P_{out}$ is memorized by the digital computer. Second, we measure the output for input combination $\{\pi/2, \pi/4, \pi/4, \dots \pi/4\}$. This input phase combination is equivalent to logic state $\left\{1, \frac{|0\rangle + |1\rangle}{\sqrt{2}}, \frac{|0\rangle + |1\rangle}{\sqrt{2}}, \dots \frac{|0\rangle + |1\rangle}{\sqrt{2}}\right\}$. Then, we compare the output powers for the two measurements. The "right" phase always provides a larger output. The comparison procedure is the following. The reference value $P_{ref}$ is setup to $P_{out}$ measured for $\left\{0, \frac{|0\rangle + |1\rangle}{\sqrt{2}}, \frac{|0\rangle + |1\rangle}{\sqrt{2}}, \dots \frac{|0\rangle + |1\rangle}{\sqrt{2}}\right\}$. The right phase for input $i = 1$ is $\pi/2$ (i.e., logic 1) if $P_{out}$ $\left\{1, \frac{|0\rangle + |1\rangle}{\sqrt{2}}, \frac{|0\rangle + |1\rangle}{\sqrt{2}}, \dots \frac{|0\rangle + |1\rangle}{\sqrt{2}}\right\}$ is larger than $P_{out}$ $\left\{0, \frac{|0\rangle + |1\rangle}{\sqrt{2}}, \frac{|0\rangle + |1\rangle}{\sqrt{2}}, \dots \frac{|0\rangle + |1\rangle}{\sqrt{2}}\right\}$. The right phase is $0\pi$ (i.e., logic 0) otherwise.

In step 2, the phase shifter at input $i = 1$ is fixed to the just found value (see step 1). All inputs from $i = 3$ to $i = n$ are in the superposition of states. We make two measurements with two possible phases for input 2. The one with a larger output power corresponds to the right phase. The procedure is repeated $n$ times until the complete input phase combination is found. Finally, one can apply the obtained phase combination with $P_{ref} = n^2 P_0$ to verify the answer (i.e., to confirm that the logic output is 1).

The search procedure for Oracle-C with seven inputs $n = 7$ is illustrated in Fig.2. The interferometer has the following set of phase shifters $\Delta_i\{\pi/4, -\pi/4, -\pi/4, \pi/4, \pi/4\}$. There is just one input phase combination $\{0\pi, \pi/2, \pi/2, 0\pi, 0\pi\}$ which results in the constructive interference at the output. This phase combination was found in 7 steps (14 measurements) according to the above described procedure. The three tables in Fig. 2(a) show the input phase combinations for the first three steps. The numerical values for $P_{out}$ and $P_{ref}$ for all steps are summarized in the Method section. Fig.2(b) shows the evolution of the output signal $y_{out}$ for the seven consequent steps.

$y_{out} = A_m \cdot \sin(\omega t + \phi_{out})$,

where $A_m$ is the maximum amplitude and $\phi_{out}$ is the phase of the output signal. It is convenient to express $y_{out}$ as a vector in a polar form. The red vector in Fig.2(c) depicts the output signal corresponding to the searched phase combination. It corresponds to the constructive wave interference when all $n$ input waves reach the output in-phase: $\phi_{out} = \pi/4$, $A_m = nA_0$. The black and blue markers in Fig.2(c) show the evolution of the output signal during the measurements. The black marker corresponds to the "true" phase input while the blue marker corresponds to the "false" input phase. There are seven black and blue markers corresponding to the results of 14 consequent measurements. The phase combination "true" + superposition appears closer to the correct result (i.e., the red vector) compared to the "false" + superposition combination at each step. The seventh black marker coincides with the correct phase combination.

The described above procedure is based on the comparison between the "true" + superposition and "false" + superposition phase combinations. The "true" + superposition phase combination always provides a larger output power as it closer to the constructive interference. There is no exponential overhead in terms of number of devices or power. The required accuracy of measurements does scale exponentially with the number of inputs. The difference in the output power between the "true" + superposition and "false" + superposition combinations depend on the set of phase shifters $\Delta_i$. It attains its minimum (i.e. the worst scenario) when $n - 1$ shifters are the same (e.g., all $+\pi/4$ or all $-\pi/4$) and only one phase shifter is different from the others. In Fig. 2(c), the difference is plotted as a function of the number of inputs. The red, blue and black curves depict the normalized difference between the "true" and "false" phase combinations for $n = 7$, $n = 30$, and $n = 100$. In all three cases, the first $n - 1$ phase shifters are set to $+\pi/4$ and only the last one is set to $-\pi/4$. Though the difference between the "true" and "false" phase combinations does increase with the number of inputs $n$, it appears on the acceptable level (i.e., 0.0001) even for $n = 100$. To verify the result of the search procedure for the seven-input Oracle-C, all phase combinations were checked one by one. The graph in Fig.2(d) shows $A_{out}$ normalized to $A_0$ for $2^7 = 128$ input phase combinations. Indeed, the phase combination found in 14 measurements provides the maximum output corresponding to the constructive wave interference.

The search algorithm was performed for $n = 100$ as well. It was found an input combination leading to the output logic 1 for the given Oracle-C structure. However, it is not practically possible to verify if this is the only result as it would take $2^{100}$ operations to check one by one all possible input phase combinations.

The superposition approach can be extended to a multivalued case. In this scenario, multiple logic states (e.g., $0,1,2,3, \dots m$) can be encoded into different phases (e.g., $\phi(0), \phi(1), \phi(2) \dots \phi(m)$ ). In the next example, we demonstrate a search procedure for finding the only input (i.e., phase combination) leading to the constructive wave interference out of $m^n$ possible combinations.

## Example 2: Search through an Oracle-C with multivalued inputs.

Let us consider a three-input $n = 3$ Oracle-C with eight possible logic states at each input $m = 8$ . The input logic states are encoded into eight phases $\left\{0\pi, \frac{\pi}{14}, \frac{2\pi}{14}, \frac{3\pi}{14}, \frac{4\pi}{14}, \frac{5\pi}{14}, \frac{6\pi}{14}, \pi/2\right\}$. The phase shifter $\Delta_i$ in the Oracle-C may be any of $\left\{0\pi, \frac{\pi}{14}, \frac{2\pi}{14}, \frac{3\pi}{14}, \frac{4\pi}{14}, \frac{5\pi}{14}, \frac{6\pi}{14}, \pi/2\right\}$ excluding only the same values for all three shifters (i.e., there are no combinations $\Delta_1 = \Delta_2 = \Delta_3$). Thus, Oracle-C is constructed in such a way that only one input phase combination results in the constructive interference (logic output =1 at $P_{ref} = 3P_0$). The task is to find this phase combination in the minimum number of steps.

The search procedure is illustrated in Fig.3. The whole number of possible phase combinations constitutes a cube in a 3-dimentional space as depicted in Fig.3(a). The $x, y,$ and $z$ axes correspond to the three input phases Phase1, Phase2, and Phase3, respectively. There are eight possible values for each phase. The total number of possible phase combination is $8^3 = 512$. Instead of checking one by one all these combinations, we divide the values on each axis on two halves and apply wave superposition to check phase combinations in each segment. Similar to the Example 1, the output of the phase segment with the true phase combination is always larger compared to the other segments.

In step 1, we divide the whole phase space into segments. There are eight phase segments for a 3-dimensional space. For example, Segment 1 in Fig.3(a) includes all input combinations with $\phi_1 \in \left\{0\pi, \frac{\pi}{14}, \frac{2\pi}{14}, \frac{3\pi}{14}\right\}, \phi_2 \in \left\{0\pi, \frac{\pi}{14}, \frac{2\pi}{14}, \frac{3\pi}{14}\right\}$, and $\phi_3 \in \left\{0\pi, \frac{\pi}{14}, \frac{2\pi}{14}, \frac{3\pi}{14}\right\}$. Segment 2 includes all input combinations with $\phi_1 \in \left\{0\pi, \frac{\pi}{14}, \frac{2\pi}{14}, \frac{3\pi}{14}\right\}, \phi_2 \in \left\{0\pi, \frac{\pi}{14}, \frac{2\pi}{14}, \frac{3\pi}{14}\right\}$, and $\phi_3 \in \left\{\frac{4\pi}{14}, \frac{5\pi}{14}, \frac{6\pi}{14}, \frac{\pi}{2}\right\}$. Next, we provide eight measurements with the phase combinations corresponding to the superposition of states in each segment. The phase combination for segment 1 is $\left\{\frac{\pi}{8}, \frac{\pi}{8}, \frac{\pi}{8}\right\}$. The phase combination for segment 2 is $\left\{\frac{3\pi}{8}, \frac{\pi}{8}, \frac{\pi}{8}\right\}$ and so on. The results of the measurements are summarized in the Tables in Fig.4. The first table shows the output amplitudes for the eight phase combinations. The largest amplitude of 2.84675 $A_0$, where $A_0$ is the output of a single input, is detected for Segment 2. The output amplitude is larger for the segment containing the searched phase combination. The reason is the same as illustrated in Fig.3. The superposition of waves in the right segment is closer to the constructive superposition.

In Step 2, we take phase segment 2 with provides the maximum output, divide it into the eight sub-segments and provide measurements for each sub-segment. The results of numerical simulations are summarized in Table 2. The maximum output amplitude corresponds to the phase segments containing the following phases $\phi_1 \in \left\{ \frac{2\pi}{14}, \frac{3\pi}{14} \right\}, \phi_2 \in \left\{ 0\pi, \frac{\pi}{14} \right\}$, and $\phi_3 \in \left\{ \frac{6\pi}{14}, \frac{7\pi}{14} \right\}$. There are eight phase combinations left out of the 512 possible. In the final Step, we check one by one all the remaining phase combinations. The maximum output amplitude of $3.0\ A_0$ corresponding to the constructive interference occurs for the input phase combination $\left\{ \frac{\pi}{7}, 0\pi, \frac{\pi}{2} \right\}$. One can check that only this input phase combinations provides the constructive output interference for $\left\{ \Delta_1 = \frac{5\pi}{14};\ \Delta_2 = \frac{\pi}{2};\ \Delta_3 = 0\pi \right\}$. The search was accomplished in 24 queries. It would take 256 queries in average for a digital computer. In general, the advantage of the superposition technique using phase space division over a classical digital machine is $\mathcal{O}(\sqrt[n]{m})$, where $n$ is the number of inputs (i.e. the dimension of the phase space), and $m$ is the number of states per each input (i.e. the number of phases per input).

**Example 3: Experimental data: search through a magnetic database using spin wave superposition.**

In order to validate the practical value of the described search procedures, we present experimental data on magnetic database search using spin wave superposition. Data centers based on magnetic storage technology have proved to be the core platforms for cloud computing and Big Data storage [12, 13]. It has already stimulated a search for novel and efficient methods for parallel magnetic bit read-out (e.g., multihead, multitrack magnetic memory [14]). In this part, we implement the search algorithm described in the previous examples to a magnetic database. In our preceding works [15, 16], we have developed magnonic holographic memory (MHM) aimed at exploiting spin waves for parallel read-in and read-out. Spin waves—or magnons, the quanta of spin waves—represent eigen excitations of the electron spin subsystem in magnetically ordered media and are observed in ferro- and ferrimagnets, as well as in antiferromagnets [17]. The interaction between the magnetic bits and propagating spin waves is the base of MHM operation. The examples of working prototypes are reported in Refs. [18]. The schematic of a MHM device is shown in Fig.5. It is a multi-input interferometer with a mesh of magnetic waveguides inside. Input spin waves are excited by the set of micro antennas placed on the top of waveguides (i.e. # 1-6). The output is the inductive voltage detected by the output antenna (i.e. # 7) produced by the interfering spin waves. The core of the MHM device is a mesh of magnetic waveguides made of material with low spin wave damping (e.g. $Y_3Fe_2(FeO_4)_3$ (YIG)). There are magnets (e.g., Co) placed on the top of the waveguides. These are memory elements where information is encoded in the magnetization direction. The details of the structure preparation and measurement techniques can be found in Ref. [19]. Spin waves propagating from different inputs to the output accumulate different phase shifts $\Delta_i$ which depend on the configuration of magnets in the mesh. The phase difference between the waves does not exceed $\pi/2$. The set of attenuators was used to equalize the amplitudes of the spin waves at the output port (i.e. $\sigma_i$ are the same for all inputs). We consider a linear spin wave propagation at low input power (i.e., $\sigma_i$ and $\Delta_i$ do not depend on the wave amplitude). Overall, the input-output correlation of the MHM device is well described by Oracle-C as shown in Fig.1. In our experiment, we use five input antennas (i.e., marked #1-5 in Fig. 5) to provide input information, antenna #6 to provide a reference signal with constant phase, and antenna #7 to pick up the inductive voltage. Four distinct phases $0^0, 7^0, 14^0$, and $21^0$ are used for each antenna. There are $4^5 = 1024$ possible phase combinations. The task is to apply the search algorithm as described in example 2 and find phase combination resulting in the maximum output voltage.

The ensemble of all phase combinations constitutes a cube in a 5D space. Unfortunately, it cannot be visualized as in Fig.3. In the first step, the whole phase space is divided in $2^5=32$ sub-segments, where phases for each antenna are grouped in two halves: $(0^0, 7^0)$ and $(14^0, 21^0)$. We apply wave superposition in each of the segments (e.g. $4^0$ or $8^0$ for each antenna) and detect the inductive voltage. The Tables with experimental data are shown in Fig.5(c). In step 2, we consider only one sub-segment which provides the maximum output voltage. There are 32 possible phase combinations left which are checked one by one. There are found two phase combinations $(21^0, 0^0, 0^0, 0^0, 21^0)$ and $(21^0, 7^0, 7^0, 0^0, 21^0)$ which provide the highest output voltage. In order to verify these results of the superposition-based search, we accomplished test measurements taking one by one all 1024 phase combinations. It took about two months to complete the test. The results are shown in Fig.5(b). The inductive voltage (vertical scale) in mV is shown for all phase combinations (horizontal scale). The phase combinations are numbered on as follows (e.g., $(0^0, 0^0, 0^0, 0^0, 0^0)$=1, $(0^0, 0^0, 0^0, 0^0, 7^0)$=2, ... $(21^0, 21^0, 21^0, 21^0, 21^0)$=1024. There are two phase combinations $(21^0, 0^0, 0^0, 0^0, 21^0)$ (i.e., # 515 in the graph) and $(21^0, 7^0, 7^0, 0^0, 21^0)$ (i.e., # 755 in the graph) which provide 0.9501 mV and 0.9507 mV inductive voltage, respectively. Thus, the results of one-by-one measurements confirmed the results of the superposition-based search. It took 64 measurements with Oracle-C using superposition instead of 1024 subsequent measurements. Overall, Examples 1-3 show an intriguing possibility of exploiting classical wave superposition for database search speedup.

It is worth to mention the capabilities of the classical wave-based approach for prime factorization. Though this approach cannot compete with true quantum algorithms in efficiency as it does not exploit quantum entanglement, it still may provide a fundamental advantage over the digital computers.

**Example 4: Period finding using classical wave superposition.**

Period finding is the key part of the Shor's prime factorization algorithm [20]. The algorithm includes two parts: classical and quantum. The classical part accomplished on a general type computer is used for calculating $f(k) = m^k mod(N)$ function, where $N$ is the number to be factorized, $m$ is the almost randomly chosen number. The quantum part is the period funding subroutine aimed to find the period r of the function $f(k)$. As the period is found, the classical computer checks the greatest common divisors $(gcd)$: $gcd(m^{r/2} + 1, N)$ and $gcd(m^{r/2} - 1, N)$. At least one of $gcd$ is a nontrivial factor of $N$. The period finding is the most challenging part for a classical digital computer. For instance, let us consider a sequence:

010110111011110010011001110001110101101110111100100110011100011101011011101111100100

It consists of zeros and ones. In the most naïve and time consuming approach, one would have to check the period at every repeating zero or one to find $r = 32$. Shor developed a polynomial-time quantum algorithm exploiting quantum superposition and quantum Fourier transform to speedup the period funding part, which provides a fundamental advantage over any type of digital-type computers [20].

Period funding can be also efficiently accomplished using classical wave superposition. Let us consider a sequence of numbers $f(k)$ as a superposition of waves with phases $[f(k)/N]/\pi$. As an example, we consider $N = 3 \times 5 \times 17 = 255$. We take $m = 13$ and calculate a sequence of $f(k) = 13^k mod(255)$. The calculated numbers are converted into waves with phases $\phi(k) = [f(k)/N] \times \pi$. In Fig.6(a), it is shown *the phase* of the wave superposition as a function of $k$. The phase converges to some value (i.e. $1/3\pi$ in the given example) as $k$ increases. This phase $1/3\pi$ is nothing but the phase of wave superposition in one period. To find the period, one has to find the first $k$ with this phase. The inset in Fig.6(a) shows the enlarged part of the plot. The phase of the superposition $1/3\pi$ appears at $k = 4$, which is the period of the given modular sequence. The rest of the prime factorization is trivial. Calculating $gcd(13^{4/2} + 1,255) = 17$ and $gcd(13^{4/2} - 1,255) = 7$ gives one (i.e. 17) nontrivial factor of 255.

In Fig.6(b), we plotted the results of numerical simulations for $N = 3 \times 5 \times 11 \times 17 = 2805$ and $m = 13$. As in the previous example, the sequence of the modular function is converted into a wave superposition with $\phi(k) = [f(k)/N] \times \pi$. The phase converges to $0.41657\pi$ at large $k$. To find the period of the sequence, one takes the first $k$ with this phase. The inset to Fig6(b) shows the enlarged plot where the phase of the wave superposition is $0.41657\pi$ at $k = 20$, which is the period of the given modular sequence. We want to stress that the described procedure is universal and can be applied for any periodic function. The phase of the superposition does converge to the period value regardless of particular numbers (phases) in the sequences. This approach also provides a fundamental advantage over the classical one by reducing the number of calculations. In our preceding work [21], we used MHM device to factorize number 15.

**Discussion**

There are several observations we want to make based on the presented examples.

(i)     Classical wave-based devices may provide a fundamental speedup in database search compared to classical digital computers. For instance, Example 1 shows the possibility to find one of the $2^n - 2$ input combinations in $2n$ steps. Examples 2 and 3 demonstrate the feasibility of $\sqrt[z]{m}$ speedup, where z is the dimensionality of the phase space (i.e., the number of independent wave inputs) and $m$ is the number of phases per input. There may be a number of other examples showing the extreme capabilities of wave superposition for parallel database search.

(ii)    The speedup does not come with an exponential resource overhead. The number of devices as well as the energy of operation scales linearly with the number of inputs $n$. There is a tradeoff between the accuracy of measurements and the number of input ports as shown in Fig.2(C). However, the increase in accuracy (e.g. $10^{-4}$ for $n = 100$) is far from any theoretical/practical limits to prevent the utilization of classical wave superposition.

(iii)  The presented examples are not universal but can be applied to a special type of databases (i.e. with one absolute maximum). A detailed discussion on other types of databases, the possibility of extending the described search procedures, etc. is beyond the scope of this work.

(iv)  All of the present examples are based on the phase information coding, where logic states are related to the phases of the wave signals. The use of phase possesses certain advantages for parallel data processing (i.e. state superposition).

(v)  Besides database search, classical wave-based devices can be exploited in special type data processing (e.g. period finding). These devices may not be as efficient as quantum computers (as they operate without quantum entanglement) but provide a fundamental speedup over digital computers.

To conclude this work, we would like to refer to a recent publication by F. Arute et al., showing the Quantum supremacy using a programmable superconducting processor [22]. In this work, the team exploited both the superposition and the quantum entanglement of 53 qubits. It takes about 200 seconds for the quantum Sycamore processor to complete the task which would take approximately 10,000 year for a state-of-the-art classical supercomputer [22]. The advantages of quantum computers are undisputed. Here, we want to turn attention to classical wave based devices, which may accomplish some problems with the same efficiency as quantum computers. Superposition of states whether quantum or classical provides a fundamental advantage over digital computers. The very first example presented in this work, shows a database search over $2^{100}$ input state combinations. It will take a bit longer than the age of the universe (13.77 billion years) to check one by one all combinations (1 combination per 1 ns). It seconds to solve it using classical wave superposition. There is a big room for computing power enhancement by utilizing phase in addition to amplitude for information encoding. It may extend the Moor's Law until quantum computers will be practically available.

**Methods**

**Device Fabrication**

The magnetic interferometer used in Example 3 is mesh of waveguides with four cross junctions made of single crystal $Y_3Fe_2(FeO_4)_3$ film. The film was grown on top of a (111) Gadolinium Gallium Garnett ($Gd_3Ga_5O_{12}$) substrate using the liquid-phase epitaxy technique. The micro-patterning was performed by laser ablation using a pulsed infrared laser ($\lambda \approx 1.03$ µm), with a pulse duration of ~256 ns. The YIG cross has the following dimension: the length of the each waveguide is 3.65 mm; the width is 650 µm; and the YIG film thickness is 3.8 µm; and saturation magnetization of $4\pi M_0 \approx 1750\ G$. There are four Π-shaped micro-antennas fabricated on the edges of the cross. Antennas were fabricated from a gold wire of thickness 24.5µm and placed directly at the top of the YIG surface.

**Measurements**

The antennas are connected to a programmable network analyzer (PNA) Keysight N5241A. Two of the antennas are used to generate two input spin waves. The inductive voltage is detected by the other two antennas. The set of attenuators (PE7087) and phase shifters (ARRA 9428A) is used to control the amplitudes and the phases of the interfering spin waves.

**Author Contributions**

A.K. conceived the experiment, provided numerical modeling and wrote the manuscript, M.B. carried out the experiments, D.G. and H.C. developed experimental setup, A.Ko. and Y.F. provided the sample for experimental work. All authors discussed the data and the results, and contributed to the manuscript preparation.

**Competing financial interests**
The authors declare no competing financial interests.

**Acknowledgement**

This work was supported in part by the INTEL CORPORATION (Award #008635, Spin Wave Computing) (Project director is Dr. D. E. Nikonov). The work of A.V. Kozhevnikov and Y.A. Filimonov is supported by the Russian Science Foundation under grant 17-19-01673.

**Figure Captions**

**Figure 1:** Schematic view of the Classical Oracle machine - Oracle-C. It consists of a multi-input interferometer, a non-linear detector, and a general-purpose digital computer. Input information is encoded into the phases of classical sinusoidal waves $\phi_i$. The output of the interferometer is a wave - the result of interference of all input waves. The output logic state is defined by comparing the output power with a reference value $P_{ref}$. The digital computer controls the input phases, the reference value $P_{ref}$, and stores the results of measurement.

**Figure 2:** (A) Illustration of the search procedure. There are shown the first three consecutive steps. At the first step all inputs from 2 to *n* are put into a superposition of states. There are two measurements to determine the "true" value of input 1. One measurement is 0> + superposition the other is 1> + superposition. The "true" input always provides the larger output power. It takes *n* steps with *2n*

measurements to find the one of the $2^n$ possible input combinations. (B) Results of numerical modeling showing the evolution of the evolution of the output signal $y_{out}$ as a vector in a polar form for the seven consequent steps. The black marker corresponds to the "true" phase input while the blue marker corresponds to the "false" input phase. The red vector corresponds to the searched phase combination. (C) Results of numerical simulations showing the output amplitude for all possible input phase combinations. The amplitude is normalized to the amplitude of a single input $A_0$. (D) Results of numerical modeling showing the minimum difference between the "true" and the "false" power output. The red, the blue and the black curves correspond to 7-, 30- and 100 input Oracle –C.

**Figure 3**:  Illustration of the search procedure in Oracle –C with multivalued inputs. It is considered a three input Oracle-C with eight possible phase combinations per input. The internal structure of the Oracle is setup in such a way that only one phase combination provides output 1. The task is to find this phase combination in the minimum number of steps.  In the first step, the whole space of all possible phase combinations is divided into the eight segments. The output of each of the segments is obtained using wave superposition. The segment with the "right" phase combination provides the largest output. In the second step, the search is accomplished in the segment found in step 1.  Finally, the last eight combinations are checked one by one. The right combination provides the maximum output (i.e. the constructive wave interference).

**Figure 4:** Results of numerical modeling on the search procedure in Example 2. The first three columns show the input phases and the forth column shows the output amplitude normalized to $A_0$,  where $A_0$ is the output of a single input. Tables (A), (B) and (C) show the results for the first, second, and the third search steps, respectively.

**Figure 5:**  (A) Schematics of the experimental setup. It consists of a multi-port spin wave interferometer. There are 6 (#1-#6) input and one output  (#7) antennas. Input signals are spin waves.  The spin wave excited by the same microwave source. The set of attenuators and phase shifters is to control the amplitudes and the phases of the input signals. The output is inductive voltage produced by the interfering spin waves. More details on the interferometer material structure  and measurement procedures can be found in the Method section and Supplementary materials.  (B) Experimental data. The plot shows the output (i.e. the inductive voltage in mV for all 1024 input phase combinations. The red circles depict the maximum output voltage (i.e. the phase combinations we are looking for). (C) Tables with experimental data showing the input phase combination and the output inductive voltage for step 1. The maximum output voltage corresponds to the phase segment in 5D space containing the searched phase combination(s). (D) The summary Table showing the number of quires required for a classical digital computer and Oracle-C using wave superposition.

**Figure 6:**  Results of numerical modeling illustrating the period fining procedure from Example 4. (A) The blue markers correspond to the phase of the superposition of waves with phases $[f(k)/N]/\pi$. The phase converges to some value at large $k$. The inset shows the enlarged part of the plot. The first $k$ at which the phase reaches the converged value corresponds to the period $r$. (A) $N = 255$, $m = 13$, $r = 4$. (B) $N = 2805$, $m = 13$, $r = 20$.

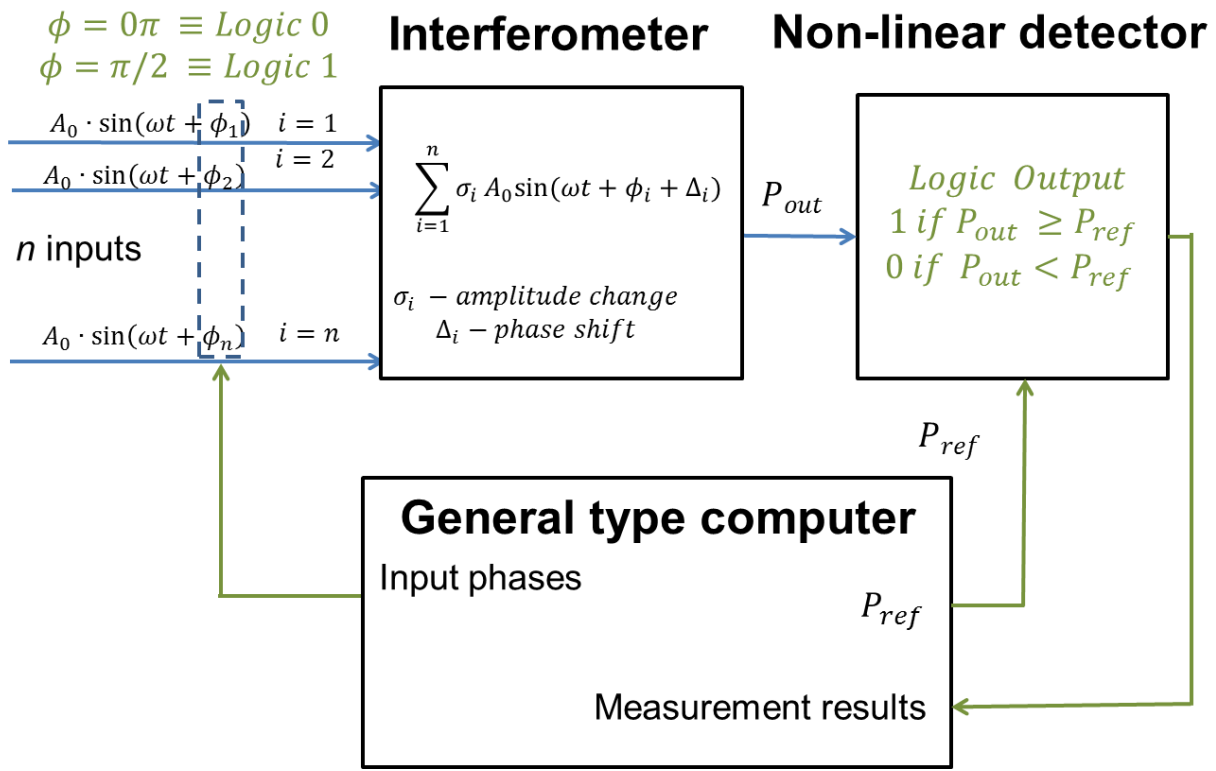

**Figure 1**

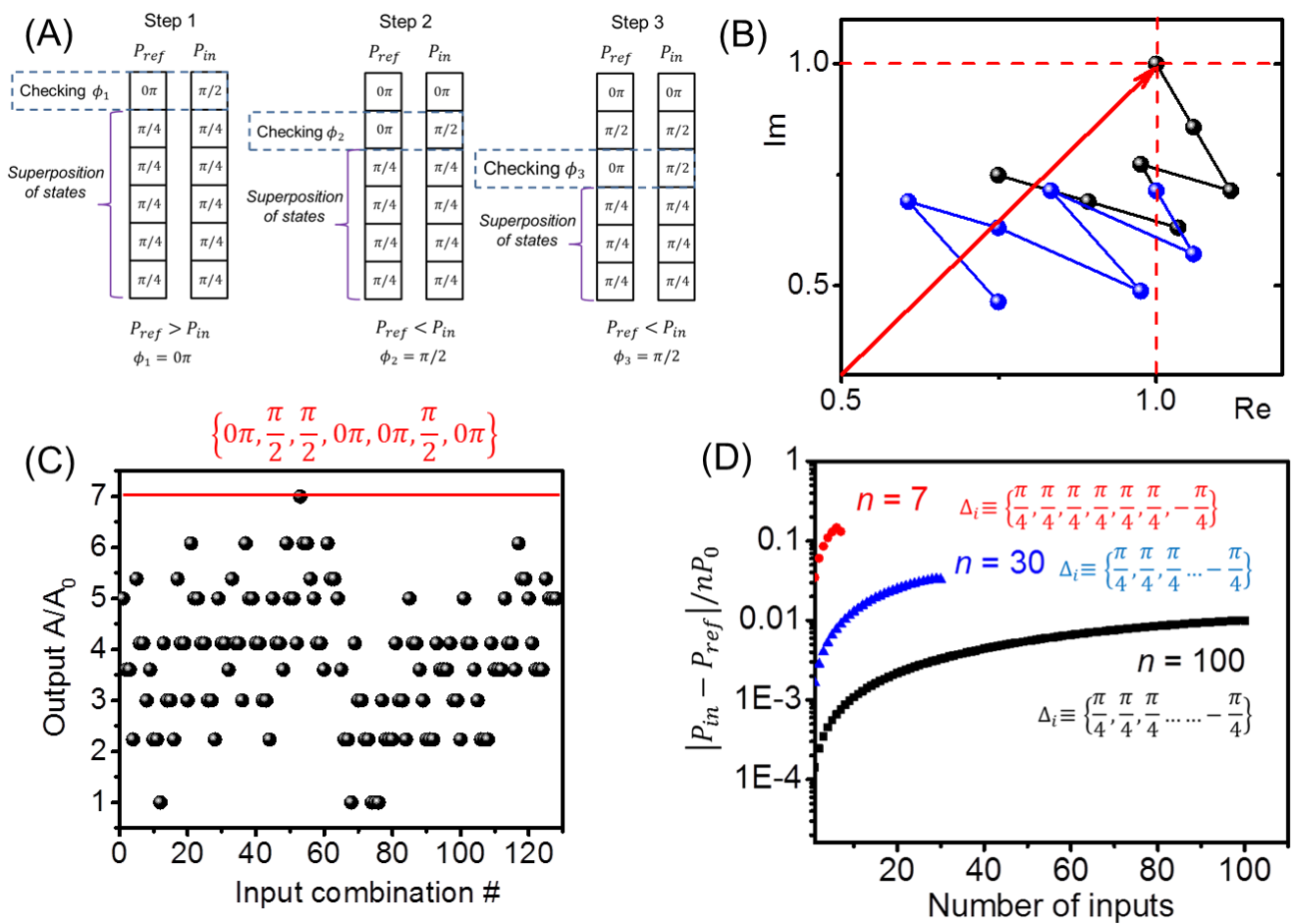

**Figure 2**

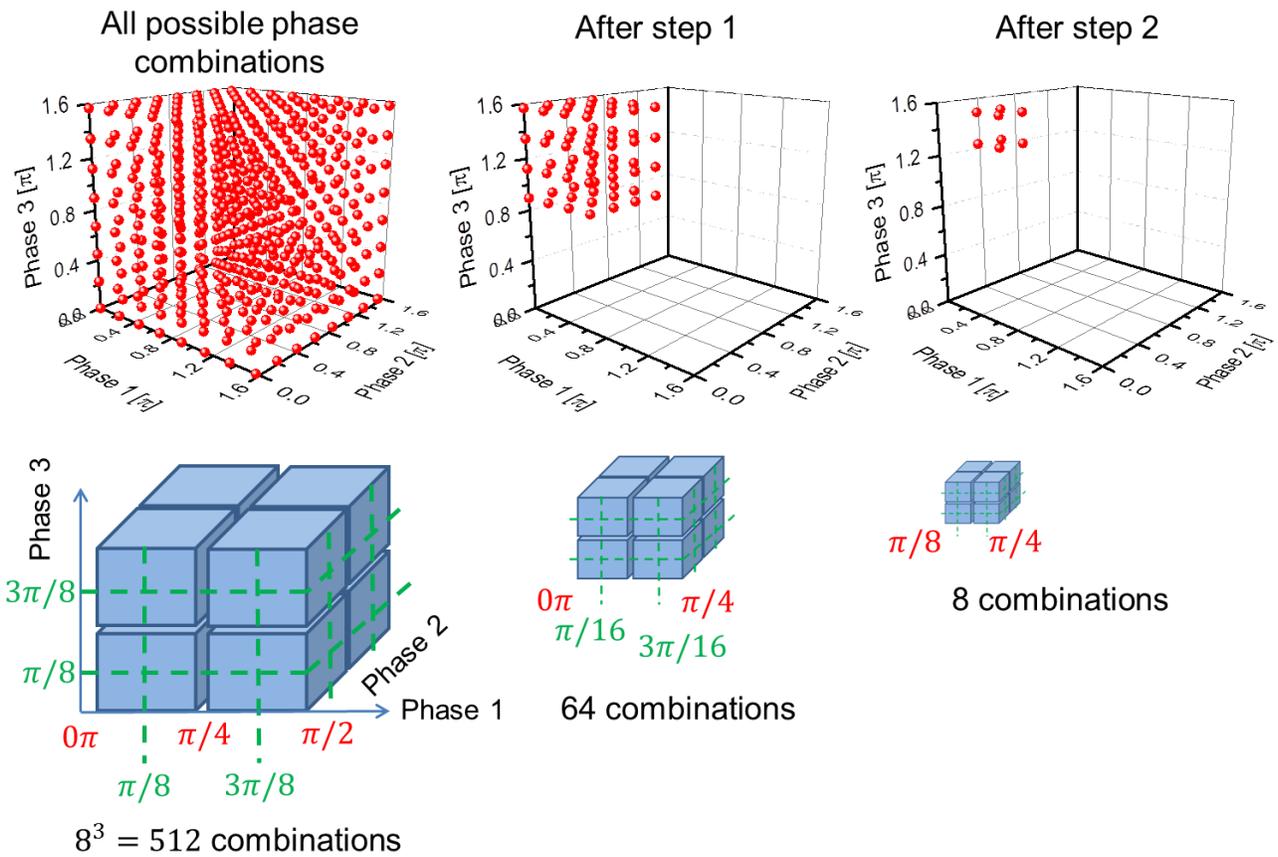

**Figure 3**

## (A) The results of the first search step

| Phase at Input 1 [π] | Phase at Input 2 [π] | Phase at Input 3 [π] | Output amplitude [$A_0$] |
|---|---|---|---|
| 1/8 | 1/8 | 1/8 | 2.38115 |
| **1/8** | **1/8** | **3/8** | **2.84675** |
| 1/8 | 3/8 | 1/8 | 1.764675 |
| 1/8 | 3/8 | 3/8 | 2.35552 |
| 3/8 | 1/8 | 1/8 | 2.05607 |
| 3/8 | 1/8 | 3/8 | 2.67766 |
| 3/8 | 3/8 | 1/8 | 1.65148 |
| 3/8 | 3/8 | 3/8 | 2.38115 |

## (B) The results of the second search step

| Phase at Input 1 [π] | Phase at Input 2 [π] | Phase at Input 3 [π] | Output amplitude [$A_0$] |
|---|---|---|---|
| 1/16 | 1/16 | 5/16 | 2.84675 |
| 1/16 | 1/16 | 7/16 | 2.9405 |
| 1/16 | 3/16 | 5/16 | 2.64309 |
| 1/16 | 3/16 | 7/16 | 2.78277 |
| 3/16 | 1/16 | 5/16 | 2.81112 |
| **3/16** | **1/16** | **7/16** | **2.95506** |
| 3/16 | 3/16 | 5/16 | 2.65682 |
| 3/16 | 3/16 | 7/16 | 2.84675 |

## (C) The results of the third search step

| Phase at Input 1 [π] | Phase at Input 2 [π] | Phase at Input 3 [π] | Output amplitude [$A_0$] |
|---|---|---|---|
| 2/14 | 0 | 6/14 | 2.98324 |
| **2/14** | **0** | **7/14** | **3.0** |
| 2/14 | 1/14 | 6/14 | 2.94986 |
| 2/14 | 1/14 | 7/14 | 2.98324 |
| 3/14 | 0 | 6/14 | 2.94986 |
| 3/14 | 0 | 7/14 | 2.98324 |
| 3/14 | 1/14 | 6/14 | 2.93324 |
| 3/14 | 1/14 | 7/14 | 2.98324 |

**Figure 4**

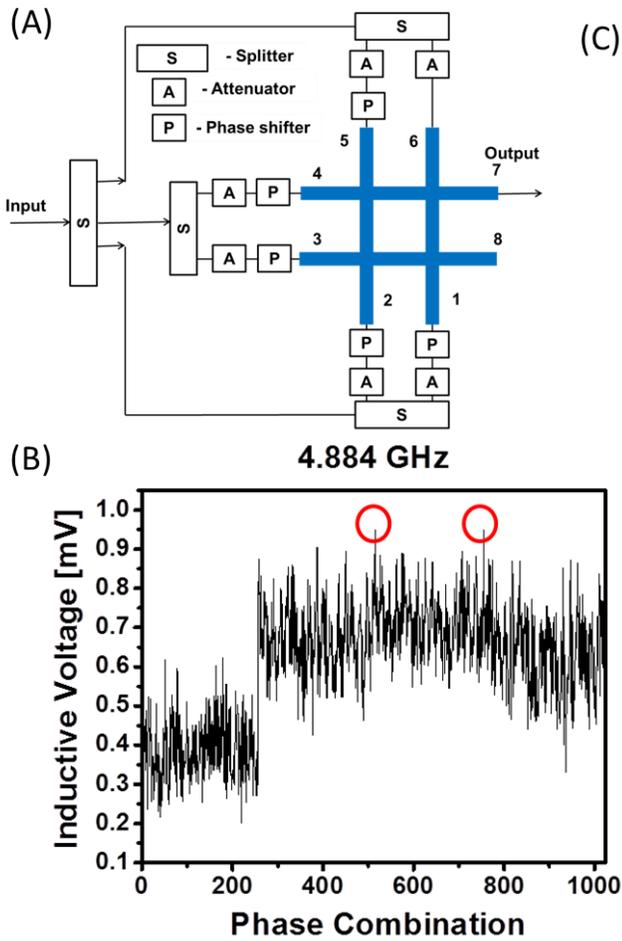

(A)

S - Splitter
A - Attenuator
P - Phase shifter

**4.884 GHz**

(B)

(C)

| Phase Combination [degrees] | Output Inductive Voltage [mV] | Phase Combination [degrees] | Output Inductive Voltage [mV] |
|---|---|---|---|
| 4, 4, 4, 4, 4 | 0.77 | 18, 4, 4, 4, 4 | 0.89 |
| 4, 4, 4, 4, 18 | 0.87 | 18, 4, 4, 4, 18 | 0.95 |
| 4, 4, 4, 18, 4 | 0.67 | 18, 4, 4, 18, 4 | 0.73 |
| 4, 4, 4, 18, 18 | 0.87 | 18, 4, 4, 18, 18 | 0.83 |
| 4, 4, 18, 4, 4 | 0.74 | 18, 4, 18, 4, 4 | 0.80 |
| 4, 4, 18, 4, 18 | 0.78 | 18, 4, 18, 4, 18 | 0.86 |
| 4, 4, 18, 18, 4 | 0.64 | 18, 4, 18, 18, 4 | 0.68 |
| 4, 4, 18, 18, 18 | 0.70 | 18, 4, 18, 18, 18 | 0.78 |
| 4, 18, 4, 4, 4 | 0.77 | 18, 18, 4, 4, 4 | 0.82 |
| 4, 18, 4, 4, 18 | 0.87 | 18, 18, 4, 4, 18 | 0.90 |
| 4, 18, 4, 18, 4 | 0.72 | 18, 18, 4, 18, 4 | 0.78 |
| 4, 18, 4, 18, 18 | 0.74 | 18, 18, 4, 18, 18 | 0.86 |
| 4, 18, 18, 4, 4 | 0.75 | 18, 18, 18, 4, 4 | 0.74 |
| 4, 18, 18, 4, 18 | 0.85 | 18, 18, 18, 4, 18 | 0.87 |
| 4, 18, 18, 18, 4 | 0.60 | 18, 18, 18, 18, 4 | 0.67 |
| 4, 18, 18, 18, 18 | 0.72 | 18, 18, 18, 18, 18 | 0.82 |

(D)

| | Classical Computer | Oracle -C |
|---|---|---|
| Number of queries | 1024 | 64 |

**Figure 5**

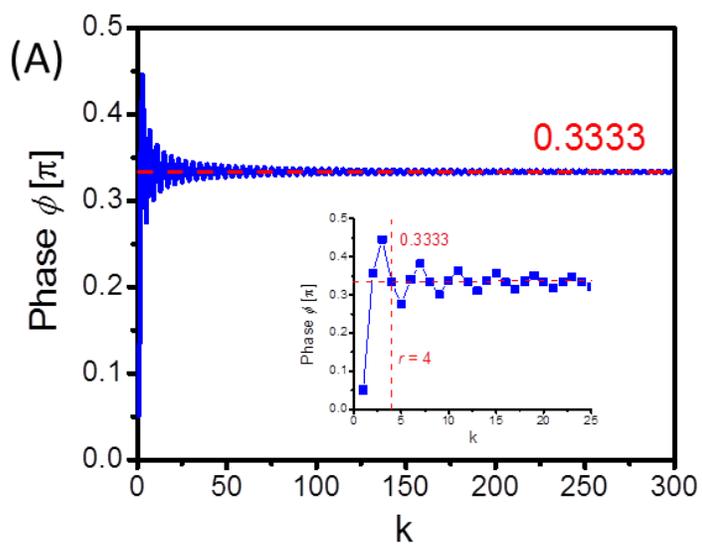 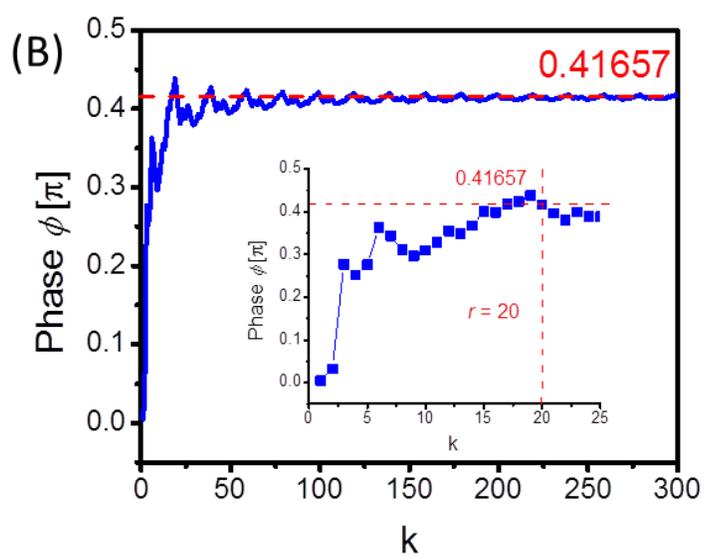

Figure 6